# Hornet Node and the Hornet DSL:

*A Minimal, Executable Specification for Bitcoin Consensus*

**Toby Sharp**

v1.2
23 September 2025
[toby@hornetnode.org](mailto:toby@hornetnode.org)

## 1. Abstract

Bitcoin's consensus rules [1] are encoded in the implementation of its reference client [2]: "The code is the spec." Yet this code is unsuitable for formal verification due to side effects, mutable state, concurrency, and legacy design.

A standalone formal specification would enable verification both across versions of the reference client and against new client implementations, strengthening decentralization by reducing the risk of consensus-splitting bugs. Yet such a specification has long been considered intractable given the complexity of Bitcoin's consensus logic (e.g. [4]).

We demonstrate a compact, executable, declarative C++ specification of Bitcoin consensus rules that syncs mainnet to tip in a few hours on a single thread. We also introduce the Hornet Domain-Specific Language (DSL) specifically designed to encode these rules unambiguously for execution, enabling formal reasoning, consensus code generation, and AI-driven adversarial testing.

Our spec-driven client Hornet Node [3] offers a modern and modular complement to the reference client. Its clear, idiomatic style makes it suitable for education, while its performance makes it ideal for experimentation. We highlight architectural contributions such as its layered design, efficient data structures, and strong separation of concerns, supported by production-quality code examples. We argue that Hornet Node and Hornet DSL together provide the first credible path toward a pure, formal, executable specification of Bitcoin consensus.

```cpp
// Performs contextual block validation, aligned with Core's ContextualCheckBlock function.
[[nodiscard]]
inline auto ValidateBlockContext(const protocol::Block& block,
                                 const int height,
                                 const AncestorTimestampsView& ancestry)
                              -> std::expected<void, BlockError> {
  static constexpr std::array ruleset = {
    // All transactions in the block MUST be final given the block height and locktime rules.
    Rule{ ValidateTransactionFinality },
    // From BIP34, the coinbase transaction's scriptSig MUST begin by pushing the block height.
    Rule{ ValidateCoinbaseHeight,       BIP::HeightInCoinbase },
    // From BIP141, the coinbase transaction MUST include a valid witness commitment
    // for blocks containing witness data.
    Rule{ ValidateWitnessCommitment,    BIP::SegWit          },
    // A block's total weight MUST NOT exceed 4,000,000 weight units.
    Rule{ ValidateBlockWeight }
  };
  BlockValidationContext context{block, height, ancestry};
  return ValidateRules<BlockError>(ruleset, context.height, context);
}
```

**Figure 1.** *Hornet Node's executable, declarative, constrained C++ implementation of Bitcoin's contextual block validation rules. These rules (with others shown elsewhere) are used today to sync and validate all mainnet headers in under 3 seconds, and all blocks in a few hours in a single thread. See text for details, and Figures 6-8 for more validation code.*

## 2. Motivation

### 2.1 The Need for Client Diversity

Bitcoin's consensus rules, first introduced in [1], are today defined only by the reference client [2]. Today the network shows ~22% of Bitcoin nodes are now Knots clients [5, 6], a derivative of Bitcoin Core [2]. While a variety of clients is an inevitable sign of healthy decentralization, we currently lack a formal specification that can be robustly implemented, tested, and ideally formally proven by a client. This presents the risk that buggy clients will experience consensus splits, provoking user confusion, media FUD, and potentially hard forks. Such risks are not theoretical: in 2013, a BerkeleyDB incompatibility in Core caused a consensus split [7].

### 2.2 Protocol Ossification vs Code Evolution

While there may be strong arguments for the ossification of the protocol, the same cannot be said for any codebase. Software must be maintained to be able to run on current hardware and operating systems, to fix bugs, and to adhere to design principles. Moreover, programming languages evolve, and each generation will have different ways to express logic. To attract talented developers of the future, the reference client should also allow for refactoring and improvement.

### 2.3 The Goals of Formal Specification

The above risks could be substantially mitigated by prioritizing a pure specification of consensus rules separate from its implementation. Such a spec would enable plaintext readability, LLM reasoning, full-coverage automated testing, and eventually formal verification. The end goal would be a formal proof that a given client is consensus-correct. This may become imperative for institutions with large bitcoin treasuries.

A declarative spec also allows precise, compact expression of BIPs as composable validation rules, facilitating analysis of their effects and enabling audit of their correctness. Likewise, rulesets can be modified to simulate hypothetical consensus forks, enabling exploration of edge cases and stress-testing of invariants in a controlled environment.

### 2.4 The Limitations of Formal Verification

While the goal of formal provability for a client is desirable, it is not possible to do formal reasoning directly against the reference client today. Theorem provers like Coq [11] work with small, constrained programs to compile and transform expression trees through a vast high-dimensional space. By contrast, Bitcoin Core is a large, entangled imperative codebase with side effects, mutable state, and concurrency.

**2.5 Previous Work**

Previous attempts at specification have been limited in scope. BitML [8] focused on reasoning about Bitcoin contracts at the transaction level, while Simplicity [9] proposed a verifiable language for scripting. Informal community documentation efforts were short-lived and never reached executable semantics.

An ongoing effort with Bitcoin Core is the creation of `libbitcoinkernel` [14], a stateful library that encapsulates consensus and validation logic while separating it from wallets, RPC, and GUI. While `libbitcoinkernel` is an incremental refactoring of existing imperative code, Hornet is designed from the outset as a declarative-style, implementation-neutral specification.

## 3. Specifying Consensus

The March 2013 chain fork [7] exposed the danger of implicit consensus rules. Nodes running v0.7 with BerkeleyDB rejected a block that exceeded the database's lock limit, while v0.8 nodes with LevelDB accepted it, producing a consensus split. The resolution was that the minority miners running v0.8 downgraded to v0.7, effectively canonizing the BerkeleyDB lock limit as a consensus rule. In subsequent releases, Bitcoin Core made this implicit rule explicit by adding validation logic to reject blocks with more than 10,000 UTXO changes.

This episode illustrates the core principle of Hornet: *the safest practice is for consensus rules to be **explicitly specified and independent** of implementation details*. What was done reactively in 2013 for one rule, Hornet seeks to do systematically for the entire protocol.

With this motivation, our approach is a declarative, executable specification of Bitcoin's consensus rules, designed under language constraints such as pure functions, explicit state transitions, immutability by default, and a strict avoidance of side effects or concurrency.

We first show how this specification drives validation in Hornet Node, our *de novo*, modern C++ client. We then introduce the Hornet DSL: a canonical, implementation-neutral domain-specific language for Bitcoin consensus, designed to be easily and unambiguously parsed, audited, and reasoned about--whether by humans, LLMs, or theorem provers.

We then outline Hornet Node's other architectural contributions and show selected code examples, concluding with a discussion of future work.

## 4. Hornet Node

Hornet Node [3] is a consensus-compatible Bitcoin client designed from the ground up to be modular, rigorous, efficient, and modern. Developed with reference to Bitcoin Core's behavior but without any copied code or external dependencies, it is an independent passion project to express the elegance of the Bitcoin protocol in idiomatic modern C++.

The consensus layer is written in a strict declarative style, with immutability and pure functions enforcing a disciplined expression of consensus rules. The broader client is implemented in idiomatic modern C++, prioritizing clarity, modularity, and performance without these constraints.

A work in progress, Hornet Node currently connects to a single peer, requests and validates mainnet headers and blocks to the tip, using consensus rules that match Bitcoin Core's behavior. It uses novel data structures for timechain data and metadata, and fully supports chain reorganizations.

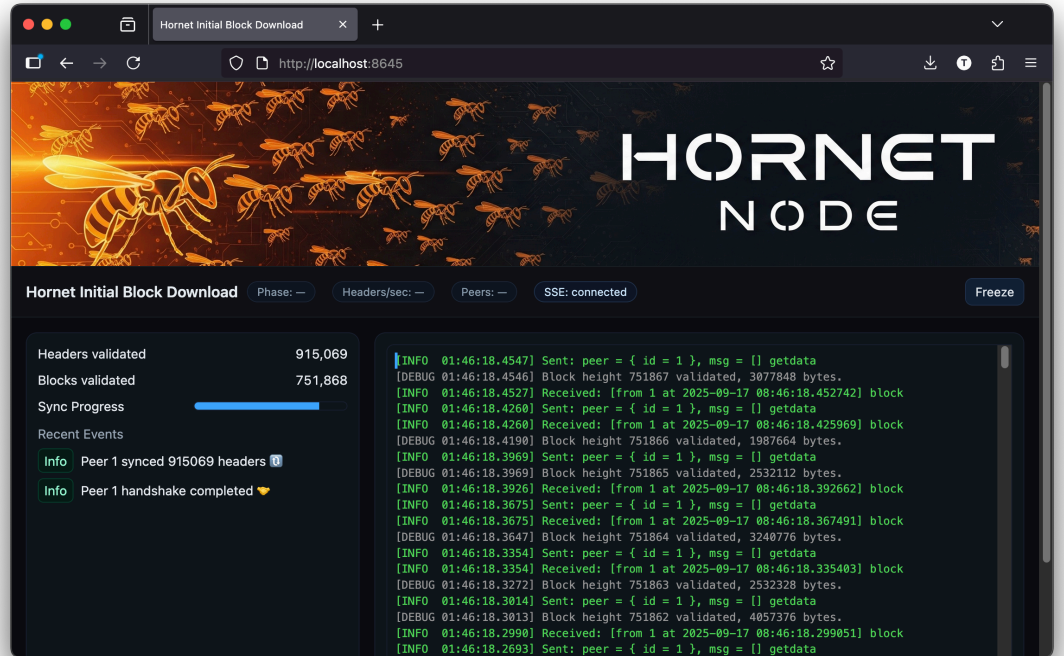

*Figure 2.* Hornet Node's interactive web UI showing Initial Block Download (IBD) syncing and validating mainnet headers and blocks against its declarative executable consensus specification.

Hornet enforces a strict one-way dependency graph (no cycles) between its hierarchical modules or *layers*, each of which is contained in its own folder and namespace. Each layer may only depend on the layers below itself. For example, `hornet::protocol` is the layer that defines protocol-specific types like `BlockHeader`. `hornet::consensus` is a layer above that may access `hornet::protocol`. But `hornet::data`, which contains data structures like the timechain itself is above `hornet::consensus`, so consensus logic has no knowledge of these implementation details. This structural discipline prevents code sprawl and limits dependency surface keeping consensus logic tight.

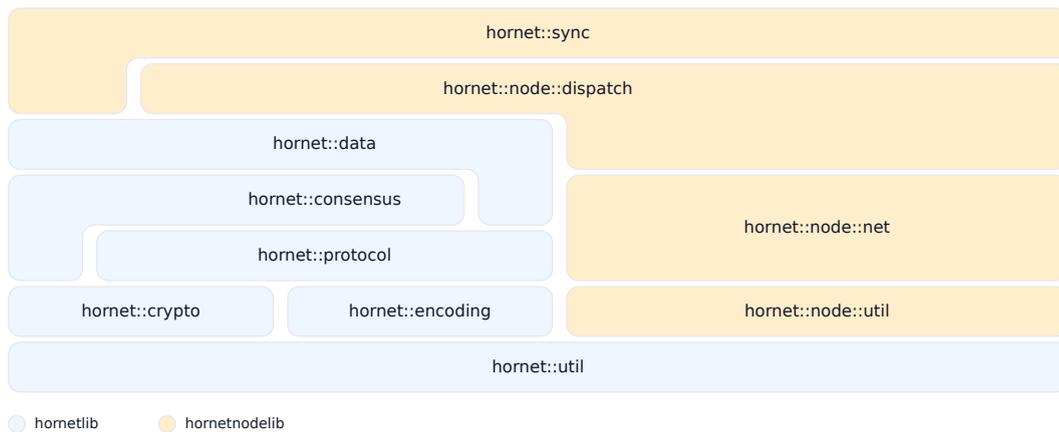

*Figure 3.* A generated layer cake diagram showing Hornet Node's layers / namespaces in stack order. Each layer may only access the layers below it in the diagram.

### 4.1 Bitcoin Script Code

Before turning to the core consensus validation code, it's worth briefly illustrating the architectural and stylistic approach that Hornet follows throughout. The following example constructs and executes a simple Bitcoin script that evaluates the expression `(21 + 21) == 42`. While this logic is not part of consensus validation, it demonstrates the principles that make the declarative specification possible: pure functional style, composable operations, and clean separation of state and behavior.

```cpp
TEST(ScriptTest, RunSimpleScript) {
    // Build a Bitcoin script to evaluate the expression (21 + 21) == 42.
    const auto script = Writer{}.PushInt(21).
                                 PushInt(21).
                                 Then(Op::Add).
                                 PushInt(42).
                                 Then(Op::Equal).Release();

    // Execute the script using the stack-based virtual machine.
    const auto result = Processor{script}.Run();

    // Assert that the script execution completed without error.
    ASSERT_TRUE(result);

    // Check the result of execution is 'true'.
    EXPECT_EQ(*result, true);
}
```

*Figure 4.* A unit test that builds and executes a Bitcoin Script to evaluate the expression `(21 + 21) == 42`, and checks that it completes execution without error and with the value `True`.

Within the script execution runtime, each opcode handler is a self-contained callable free function that receives the virtual machine state and current instruction inside a `protocol::script::runtime::Context` object. This modular style enables unit testing and formal reasoning per-opcode, avoiding the tightly coupled complexity of Core's `EvalScript`.

```cpp
// Op::PushSize1 ... Op::PushData4
static void OnPushData(const Context& context) {
  if (context.RequiresMinimal()) detail::VerifyMinimal(context.instruction);
  context.Stack().Push(context.instruction.data);
}

// Op::Add
static void OnAdd(const Context& context) {
  BinaryIntOp(context, [](int64_t a, int64_t b) { return a + b; });
}

// Op::Equal
static void OnEqual(const Context& context) {
  BinaryBitwiseOp(context, [](const auto& a, const auto& b) {
    return std::ranges::equal(a, b);
  });
}
```

***Figure 5.*** *Each opcode handler is a self-contained state-free pure function, independently testable. In contrast to Bitcoin Core's monolithic* `EvalScript` *, Hornet's handlers are modular and stateless, making script semantics easier to reason about, audit, and validate against the consensus specification.*

While script execution is not yet integrated into Hornet's block validation, the runtime shown in Figure 5 is consensus-critical and will ultimately form part of the full specification. Whether expressed in C++ or in the DSL, the structure ensures that Bitcoin Script semantics can be captured with precision, clarity, and auditability. This same design discipline--pure rules, explicit contexts, and modular handlers--carries through into Hornet's consensus-critical header and block validation, to which we now turn.

## 4.2 Header Validation

The consensus rule engine has several validation phases, the first of which is block header validation. The header consensus rules require access to state: for example, to enforce the Median Time Past rule (MTP, BIP113), a header's timestamp must be compared against the median of the past 11 ancestor timestamps. At first glance, this might suggest that consensus logic needs direct access to the header chain or node state, which would violate Hornet's strict layering model and endanger the purity of consensus rules.

Instead, Hornet introduces a minimal abstraction called `consensus::HeaderAncestryView`. This is an interface defined at the consensus layer and implemented by the data layer, allowing validation logic to query exactly the information it needs without breaking dependency boundaries. The result is a layered design that supports pure, side-effect-free rule definitions, even when those rules depend on dynamic chain state.

We now show the first part of the executable declarative C++ specification: header validation.

```cpp
// Performs header validation, aligned with Core's CheckBlockHeader and ContextualCheckBlockHeader.
[[nodiscard]] inline auto ValidateHeader(const protocol::BlockHeader& header,
                                        const model::HeaderContext& parent,
                                        const AncestorTimestampsView& view,
                                        const int64_t current_time)
                                    -> std::expected<void, HeaderError> {
  const std::array ruleset = {
    // A header MUST reference the hash of its valid parent.
    Rule{ValidatePreviousHash},
    // A header MUST satisfy the chain's target proof-of-work.
    Rule{ValidateProofOfWork},
    // A header's proof-of-work target MUST satisfy the difficulty adjustment.
    Rule{ValidateDifficultyAdjustment},
    // A header timestamp MUST be greater than the median of the prior 11.
    Rule{ValidateMedianTimePast},
    // A header timestamp MUST NOT exceed network-adjusted time plus 2 hours.
    Rule{ValidateTimestampCurrent},
    // A header version number MUST meet BIP deployment requirements.
    Rule{ValidateVersion}
  };
  const HeaderValidationContext context{header, parent, view, current_time, parent.height + 1};
  return ValidateRules<HeaderError>(ruleset, 0, context);
}
```

*Figure 6.* Hornet's executable ruleset that specifies Bitcoin's header validation.

Header validation in Hornet is defined as an ordered ruleset (Figure 6): a list of pure, composable, testable functions, each responsible for enforcing a single invariant. This function mirrors Bitcoin Core's `CheckBlockHeader()` and `ContextualCheckBlockHeader()` but expresses the logic declaratively and without side effects. Each rule operates solely on its input context, returning either a typed error code or success.

Each validation phase likewise has a ruleset comprising an ordered list of identically typed functions, each optionally tagged with a BIP for conditional operation. The implementation of all rules spans less than 50 lines, and is shown below for completeness.

```cpp
namespace hornet::consensus::rules {

using ValidateHeaderResult = std::expected<void, HeaderError>;

namespace detail {
inline bool IsVersionValidAtHeight(const int32_t version, const int height) {
  constexpr std::array<BIP, 4> kVersionExpiryToBIP = {
      BIP34, BIP34,  // v0, v1 retired with BIP34.
      BIP66,         // v2 retired with BIP66.
      BIP65          // v3 retired with BIP65.
  };
  if (version >= std::ssize(kVersionExpiryToBIP)) return true;
  const int index = std::max(0, version);
  return !IsBIPEnabledAtHeight(kVersionExpiryToBIP[index], height);
}
}  // namespace detail

// A header MUST reference the hash of its valid parent.
[[nodiscard]] inline ValidateHeaderResult ValidatePreviousHash(
    const HeaderValidationContext& context) {
  if (context.parent.hash != context.header.GetPreviousBlockHash())
    return HeaderError::ParentNotFound;
  return {};
}

// A header's 256-bit hash value MUST NOT exceed the header's proof-of-work target.
[[nodiscard]] inline ValidateHeaderResult ValidateProofOfWork(
    const HeaderValidationContext& context) {
  if (context.header.ComputeHash() > context.header.GetCompactTarget().Expand())
    return HeaderError::InvalidProofOfWork;
  return {};
}

// A header's proof-of-work target MUST satisfy the difficulty adjustment formula.
[[nodiscard]] inline ValidateHeaderResult ValidateDifficultyAdjustment(
    const HeaderValidationContext& context) {
  if (context.header.GetCompactTarget() !=
      AdjustCompactTarget(context.height, context.parent.data, context.view))
    return HeaderError::BadDifficultyTransition;
  return {};
}

// A header timestamp MUST be strictly greater than the median of its 11 ancestors' timestamps.
[[nodiscard]] inline ValidateHeaderResult ValidateMedianTimePast(
    const HeaderValidationContext& context) {
  if (context.header.GetTimestamp() <= context.view.MedianTimePast())
    return HeaderError::TimestampTooEarly;
  return {};
}

// A header timestamp MUST be less than or equal to network-adjusted time plus 2 hours.
[[nodiscard]] inline ValidateHeaderResult ValidateTimestampCurrent(
    const HeaderValidationContext& context) {
  constexpr int kTimestampTolerance = 2 * 60 * 60;
  if (context.header.GetTimestamp() > context.current_time + kTimestampTolerance)
    return HeaderError::TimestampTooLate;
  return {};
}

// A header version number MUST meet deployment requirements depending on activated BIPs.
[[nodiscard]] inline ValidateHeaderResult ValidateVersion(
    const HeaderValidationContext& context) {
  if (!detail::IsVersionValidAtHeight(context.header.GetVersion(), context.height))
    return HeaderError::BadVersion;
  return {};
}

}  // namespace hornet::consensus::rules
```

**Figure 7.** The full ruleset implementation for header consensus validation.

Let us consider `ValidateProofOfWork` in Figure 7 as an example. The comment line gives the English language plain description of the rule: *"A header's 256-bit hash value MUST NOT exceed the header's proof-of-work target."* The function first computes the SHA256^2 hash of the header, and then compares its arithmetic value against the 256-bit-expanded compact target resulting from the difficulty rule. If the hash value exceeds the target, the rule returns `InvalidProofOfWork`, otherwise it returns success. All other rules enforce their own specific constraints.

Note that all variables are declared as `const` to enforce immutability. While the validation rules delegate to methods like `BlockHeader::ComputeHash()` etc., these are immutable, deterministic, side-effect-free, and bound to protocol data only, making them effectively pure in the context of consensus specification.

The same declarative style extends beyond headers to encompass transactions and blocks. In the next section we show how these validation phases compose into a complete executable specification of Bitcoin consensus rules that is compact, rigorous, and verifiable.

## 4.3 Bitcoin Consensus As A Declarative Executable Specification

We now show the full rulesets for Bitcoin consensus header validation, block validation, and transaction validation in C++. Note that transaction validation is enforced inside the block rule `ValidateTransactions` via composability. Note also the BIP34 and BIP141 tags to restrict contextual block rules by activation height.

Remarkably, we can express the full set of Bitcoin's header, transaction, and block validation rules in fewer than 50 lines of executable idiomatic C++, where each rule encapsulates one semantic criterion of consensus. The code below is a fully executable and performant implementation of Bitcoin's validation logic.

```cpp
// Performs header validation, aligned with Core's CheckBlockHeader and ContextualCheckBlockHeader.
[[nodiscard]] inline auto ValidateHeader(const protocol::BlockHeader& header,
                                         const model::HeaderContext& parent,
                                         const AncestorTimestampsView& view,
                                         const int64_t current_time)
    -> std::expected<void, HeaderError> {
  const std::array ruleset = {
    // A header MUST reference the hash of its valid parent.
    Rule{ValidatePreviousHash},
    // A header MUST satisfy the chain's target proof-of-work.
    Rule{ValidateProofOfWork},
    // A header's proof-of-work target MUST satisfy the difficulty adjustment.
    Rule{ValidateDifficultyAdjustment},
    // A header timestamp MUST be greater than the median of the prior 11.
    Rule{ValidateMedianTimePast},
    // A header timestamp MUST NOT exceed network-adjusted time plus 2 hours.
    Rule{ValidateTimestampCurrent},
    // A header version number MUST meet BIP deployment requirements.
    Rule{ValidateVersion}
  };
  const HeaderValidationContext context{header, parent, view, current_time, parent.height + 1};
  return ValidateRules<HeaderError>(ruleset, 0, context);
}

// Performs transaction validation, aligned with Core's CheckTransaction function.
[[nodiscard]] inline auto ValidateTransaction(const protocol::TransactionConstView transaction) {
  static constexpr std::array ruleset = {
    // A transaction MUST contain at least one input.
    Rule{ValidateInputCount},
    // A transaction MUST contain at least one output.
    Rule{ValidateOutputCount},
    // A transaction's serialized size (excluding witness data) MUST NOT exceed 1,000,000 bytes.
    Rule{ValidateTransactionSize},
    // All output values MUST be non-negative, and their sum MUST NOT exceed 21,000,000 coins.
    Rule{ValidateOutputValues},
    // A transaction's inputs MUST reference distinct outpoints (no duplicates).
    Rule{ValidateUniqueInputs},
    // In a coinbase transaction, the scriptSig MUST be between 2 and 100 bytes inclusive.
    Rule{ValidateCoinbaseSignatureSize},
    // A non-coinbase transaction's inputs MUST have non-null prevout values.
    Rule{ValidateInputsPrevout}
  };
  return ValidateRules<TransactionError>(ruleset, 0, transaction);
}

// Performs non-contextual block validation, aligned with Core's CheckBlock function.
[[nodiscard]] inline auto ValidateBlockStructure(const protocol::Block& block) {
  static constexpr std::array ruleset = {
    // A block MUST contain at least one transaction.
    Rule{ValidateNonEmpty},
    // A block's Merkle root field MUST equal the Merkle root of its transaction list.
    Rule{ValidateMerkleRoot},
    // A block's serialized size (before SegWit) MUST NOT exceed 1,000,000 bytes.
    Rule{ValidateOriginalSizeLimit},
    // A block MUST contain exactly one coinbase transaction, and it MUST be the first transaction.
    Rule{ValidateCoinbase},
    // All transactions in a block MUST be valid according to transaction-level consensus rules.
    Rule{ValidateTransactions},
    // The total number of signature operations in a block MUST NOT exceed the consensus maximum.
    Rule{ValidateSignatureOps}
  };
  // clang-format on
  return ValidateRules<BlockOrTransactionError>(ruleset, 0, block);
}

// Performs contextual block validation, aligned with Core's ContextualCheckBlock function.
[[nodiscard]] inline auto ValidateBlockContext(const protocol::Block& block,
                                               const int height,
                                               const AncestorTimestampsView& ancestry) {
  static constexpr std::array ruleset = {
    // All transactions in the block MUST be final given the block height and locktime rules.
```

```
        Rule{ValidateTransactionFinality},
        // From BIP34, the coinbase transaction's scriptSig MUST begin by pushing the block height.
        Rule{ValidateCoinbaseHeight,      BIP::HeightInCoinbase },
        // From BIP141, the coinbase transaction MUST include a valid witness commitment
        // for blocks containing witness data.
        Rule{ValidateWitnessCommitment,   BIP::SegWit          },
        // A block's total weight MUST NOT exceed 4,000,000 weight units.
        Rule{ValidateBlockWeight}
    };
    const BlockValidationContext context{block, ancestry, ancestry.Length()};
    return ValidateRules<BlockError>(ruleset, context.height, context);
}
```

> ***Figure 8.*** *A complete, declarative, pure, and functioning specification for Bitcoin's header, transaction, and block validation rules in 50 code lines of idiomatic modern C++.*

These compact C++ rulesets demonstrate that Bitcoin's consensus can be expressed declaratively, without hidden state or side effects, and still execute at full performance on mainnet. Yet C++ remains a general-purpose language: it allows styles that drift from this discipline, and its semantics are not tailored to formal reasoning. To go further, we introduce Hornet DSL: a purpose-built language that encodes consensus rules unambiguously and enforces constraints by design.

## 5. Hornet DSL

### 5.1 Design goals

We have shown that Bitcoin's consensus rules can be expressed compactly and declaratively in C++. Hornet DSL takes this further: a purpose-built language that enforces purity, immutability, and composability at the syntax level, while providing natural expression with built-in knowledge of Bitcoin concepts.

Whereas C++ permits many styles, Hornet DSL constrains expression to constructs needed for pure consensus specification, making every rule's action bounded and semantically clear. The goal is an unambiguous, implementation-independent, compilable specification of Bitcoin consensus, suitable for automated testing, code generation, and formal verification.

Even without full formal methods, Hornet DSL's constrained and regular structure enables more effective analysis and LLM-based reasoning, compared to the highly imperative and stylistically diverse reference client.

### 5.2 Language features

With these goals in mind, Hornet DSL is defined to include the following features:-

- Deterministic operation (functions executed with identical inputs yield identical outputs)
- All state must be local and explicit (no globals, statics, or member functions)
- All functions execute in a single-threaded context
- All variables are immutable unless explicitly marked mutable
- All functions must return a value and have no side effects
- Types are built-in or plain structs
- Protocol structs natively define their serialization format
- A validation `rule` is a function that returns success or a typed error
- Rulesets are statically analyzable sequences of rules with uniform error types

- Validation rules are invoked with the `require` keyword, which passes the enclosing function's arguments to the referenced rule, returns early on error, or otherwise proceeds
- Validation rules may be tagged with a `@bip` annotation for selective application depending on block height
- Relevant cryptographic primitives (e.g. uint256, SHA256, secp256k1) are defined as built-in types and functions

### 5.3 Example: Contextual block validation

In this context, we present the Hornet DSL version of Figure 1:

```
// Performs contextual block validation, aligned with Core's ContextualCheckBlock function.
@rule @phase("block_context")
rule ValidateBlockContext(block: Block,
                          height: uint32,
                          past_timestamps: array<uint32, 11>)
                    -> BlockError? {

  // All transactions in the block MUST be final given the block height and locktime rules.
  require ValidateTransactionFinality
  // From BIP34, the coinbase transaction's scriptSig MUST begin by pushing the block height.
  @bip(BIP34) require ValidateCoinbaseHeight
  // From BIP141, the coinbase transaction MUST include a valid witness commitment for blocks containing witness data.
  @bip(BIP141) require ValidateWitnessCommitment
  // A block's total weight MUST NOT exceed 4,000,000 weight units.
  require ValidateBlockWeight
}
```

*Figure 9.* Designing Hornet DSL for Bitcoin consensus specification. This function is the DSL equivalent of the C++ code in Figure 1.

### 5.4 Testing, Reasoning, and Guarantees

Hornet DSL is a work in progress, iteratively informed by Hornet Node's evolving declarative C++. Once complete, the DSL will be a human- and machine-readable precise specification. An early goal at that stage will be to generate C++ that matches Hornet Node's validation pipeline. We will then be able to set up cloud-scale automated testing that continuously validates Hornet DSL backends against Hornet Node and Bitcoin Core.

We believe that it becomes highly plausible to use LLM-based models (with access to source code and compiler) in a test framework to continuously analyze the Hornet DSL and C++ specifications, assess their agreement with Bitcoin Core behavior, and generate differential adversarial test cases in a feedback-driven loop. This includes both differential testing [10] (comparing outputs across multiple implementations) and adaptive test generation (where the results of one probe inform the construction of the next).

Such targeted probing provides far more effective search of the high-dimensional space of all blocks than general fuzzing [12], leading to much stronger evidence that the specification indeed matches the Bitcoin reference client. After testing one billion blocks for potential edge case bugs, we will have simulated ~20,000 years of validated consensus behavior.

Of course, formal reasoning remains the goal for pure and mathematical proof. But this *semi-formal* approach may allow us to make quantifiable guarantees about consensus correctness. The space of all possible blocks is too vast to enumerate. Yet the space of all possible code paths in a structured specification like Hornet is very much smaller. With this approach, we believe we can reduce the probability of consensus bugs to an arbitrarily small value.

We now return to Hornet Node to describe its other key implementation contributions.

## 6. Implementation Details

Beyond its declarative consensus specification, Hornet Node is designed for conciseness, memory efficiency, and high performance. These goals make Hornet well suited for education and experimentation, and lay the groundwork for high-performance applications. In this section, we highlight three concrete examples from Hornet's implementation.

### 6.1 Protocol Messages

Hornet's protocol message system is intentionally crisp and simple. A single-threaded message loop polls peers for available reads and writes, parses messages into inbox queues, processes those messages, serializes queued outbound messages, and performs bookkeeping tasks.

Each stage of the pipeline is bound by a timeout, ensuring that a chatty or malicious peer cannot monopolize CPU time. As inbound messages are dequeued, they are dispatched using the visitor pattern [13] to all registered message subscribers.

```cpp
// Process queued inbound messages until the timeout expires.
// Using a sensible timeout prevents starvation of other duties for this thread.
void ProtocolLoop::ProcessMessages(const util::Timeout& timeout) {
  // Create a snapshot of peers and shuffle order for fairness:
  // A noisy peer may dominate this frame, but shuffling prevents systemic bias.
  const auto peers = peers_.Snapshot(/*shuffle=*/true);

  // Iterate over per-peer message inbox queues.
  for (const auto peer : peers) {
    if (peer->IsDropped() || !timeout) continue;
    auto& inbox = inboxes_[peer->GetId()];

    // Per-peer fault isolation so that one bad peer doesn't affect others.
    try {
      while (timeout && !inbox.empty()) {
        auto message = std::move(inbox.front()); inbox.pop();
        for (EventHandler* handler : event_handlers_)
          message->Notify(*handler);  // Double-dispatch via visitor pattern.
      }
    } catch (const std::exception& e) {
      // Treat all unhandled exceptions as protocol violations and drop the peer.
      peer->Drop();
    }
  }
}
```

*Figure 10. \*Hornet Node's protocol loop performs fair, bounded processing of inbound messages. Each message is dispatched to interested subscribers using the visitor pattern [13]. In contrast to Bitcoin Core's monolithic `ProcessMessage` function (~1,500 lines), Hornet separates message parsing, dispatch, and handling, keeping concerns modular and the core loop minimal.*

Subscribers implement only the message handlers they care about by overriding the appropriate virtual functions. Each handler receives a fully deserialized, strongly typed

message object:

```
void PeerNegotiator::OnMessage(const protocol::message::Ping& ping) {
  Reply<protocol::message::Pong>(ping, ping.GetNonce());
}
void PeerNegotiator::OnMessage(const protocol::message::Verack& verack) {
  AdvanceHandshake(GetPeer(verack), protocol::Handshake::Transition::ReceiveVerack);
}
```

*Figure 11.* Each message subscriber processes only the messages it cares about. Message types are dispatched with fully deserialized, strongly typed arguments. Here, the peer negotiator responds to `ping` with `pong` and to `verack` by advancing the handshake state machine.

## 6.2 Chain-Tree Data Structure

The data structure of the timechain is semantically a tree: every block has exactly one parent, except the genesis, which is the root. However, an efficient representation should recognize that the vast majority of the data is a linear chain, with ancient leaves no longer relevant, and potential forks limited to a few recent blocks.

Another observation is that, with the exception of non-pruned nodes responding to `getdata` messages and RPC requests, access to timechain elements is local: validation proceeds from genesis to tip in order, and new blocks arriving are required to build on top of some recent block. Therefore a hash map that indexes all historical blocks is not necessarily required in the minimal case.

We leverage these observations to design a novel data structure that we call the chain-tree. A chain-tree is a composition of a linear array (the main chain) and a forest of forks. Every element in the timechain's semantic tree is either in the linear array (if it's on the main chain), or in the forest branching off that chain. Each element in the forest is either an internal node that points to its parent in the forest, or it is a tree root that references its parent by index into the array.

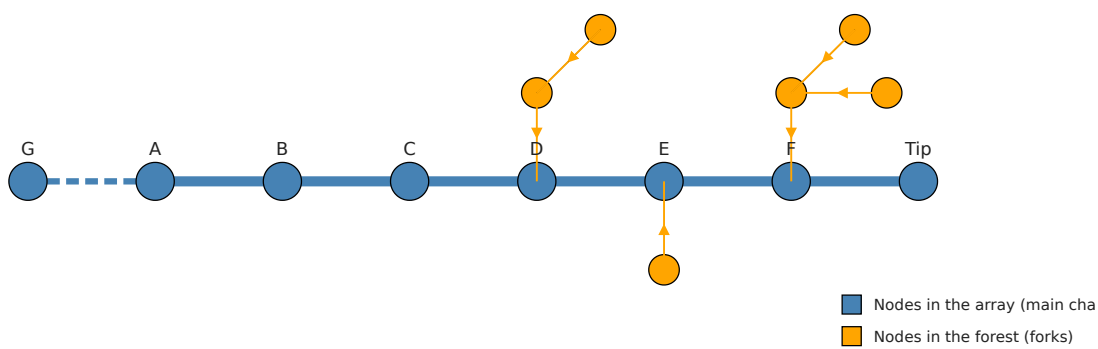

*Figure 12.* Hornet Node's chain-tree data structure is optimized for memory efficiency and fast access of blockchain data by using a flat array for the main chain and a lightweight forest for all forks.

This chain-tree composite data structure may be treated in its public API like a tree, but internally it uses its structural knowledge to access elements through the array in the vast majority of cases.

The chain-tree representation has benefits for both memory and performance. By avoiding a large hash map or node pointers, the memory requirement is barely more than the size of its linear array (~69 MB for all mainnet headers). Moreover, almost all accesses are direct memory reads. But even in the minority case, hash lookups are O(1) on a tiny map.

```cpp
// Locator is used to resolve an element *either* by height in the main branch,
// *or* by hash in the forest of forks. This locator is transferable between
// ChainTree instances that have the exact same structure and hashes.
using Locator = std::variant<int, protocol::Hash>;

// ChainTree is a data structure that represents a deep, narrow tree by using a hybrid layout: a
// linear array for the main chain combined with a forest for forks near the
// tip. This design is well suited to timechain data, where the vast majority of history is linear,
// but recent forks and reorgs must be handled too. The structure avoids a full hash map and
// minimizes per-node memory usage, making it efficient in both memory use and lookup performance.
template <
    // The data type to store at each position in the main chain and forest.
    typename TData,
    // A richer context type that includes TData, the header hash, block height, and optionally
    // additional metadata (e.g. cumulative work). This is stored at each node of the forest.
    typename TContext = ContextWrapper<TData>>
class ChainTree {
 public:
    bool Empty() const;
    int Size() const;
    ConstFindResult ChainTip() const;
    ConstIterator Find(const Locator& locator) const;
    ConstFindResult FindInTipOrForks(const protocol::Hash& hash) const;
    const TData& GetAncestorAtHeight(ConstIterator tip, int height) const;
    ...
    Iterator Add(ConstIterator parent, const TContext& context);
    void EraseBranch(Iterator root);
    // This method performs a chain reorg, i.e. it walks from the given tip node in the forest up
    // to its ancestor fork point in the chain, then swaps branches between the chain and the forest.
    // Returns the updated iterator for the now-invalidated tip.
    PromoteResult PromoteBranch(Iterator tip, std::span<const protocol::Hash> old_chain_hashes);
    ...
 private:
  std::vector<TData> chain_;
  HashedTree<NodeData> forest_;
};
```

**Figure 13.** Selected API methods from `ChainTree<T>`, a hybrid array + forest data structure. Iterators traverse upward along parent links towards the root (genesis).

An important feature of the `ChainTree<T>` class is the ability to promote a branch from the forest to become the new main chain. When this occurs, the segment of the array from the fork point onward is moved into a new tree in the forest. The invariant is preserved: every node resides either in the array or in the forest. This symmetry provides a natural mechanism for handling chain reorganizations and lays the groundwork for the sidecar design described next.

### 6.3 Metadata Sidecars

Beyond the obvious 80-byte block headers, we also need to store various metadata that applies to each node of the timechain. One good example is the block validation status that starts unset and must be updated per-block as validation occurs. The obvious temptation would be to stuff the validation status and any other metadata fields into a per-block structure with the header. However, validation status, storage location, and various other metadata are all application-level concerns, not intrinsic to the timechain itself. Separation of concerns dictates we must find another solution, but this must remain amenable to chain reorganization.

It would be possible to create a large hash map from block hashes to metadata, but this is hardly efficient. Each metadata field would require either its own hash map, or inclusion in an omnibus structure with unrelated fields. And the hash map itself consumes a disproportionate amount of memory. In Hornet Node, we avoid all such maps.

Instead we build on the `ChainTree` structure described above to create a `Sidecar<T>` type that mirrors the structure of the headers `ChainTree`. In other words, a `Sidecar<T>` has a `ChainTree<T>` to store its elements and this internal `ChainTree` is maintained in lockstep with the header chain. When a chain reorganization occurs in the headers due to a fork becoming the heaviest tip, the reorganization data is propagated to all metadata sidecars registered with the timechain, keeping their structure in sync with the header chain. Such reorganizations are small and relatively rare but need to be handled precisely.

Given an identical `ChainTree` structure between the headers and the sidecars, the same locator, derived from a stable key of `(height, hash)`, can be used to access data and metadata across structures. As before, if the item is in the main chain, no lookup is required, and instead a direct random access is performed. The necessary locking of the timechain state during a sidecar update is handled transparently.

```
// Defines our local metadata type.
enum class BlockValidationStatus { Unchecked, ... };

// Creates a keyframe sidecar to store validation status and registers it with the timechain.
auto validation_status = data::KeyframeBinding<BlockValidationStatus>::Create(timechain_);

// Locks the timechain and sidecars mutex, locates the element in the header chain-tree (direct
// access if it's on the main chain), sets the equivalent value in the sidecar, and unlocks.
validation_status.Set(height, hash, BlockValidationStatus::StructureValid);
```

**Figure 14.** *We show the creation of a custom chain metadata field, whose storage and lifetime are decoupled from the timechain, but with identical structure. This allows us to set per-block typed values efficiently, while respecting separation of concerns.*

This design allows metadata sidecars to live in the application layer but receive their structure from the timechain. Almost all accesses are in practice integer offsets which makes memory usage minimal and lookups fast.

In fact, the example shown above uses a `KeyframeSidecar<T>`, a special case of `Sidecar<T>` optimized for piecewise-constant metadata. The observation is that block validation proceeds linearly, so there are typically only one or two *changes* of status in the entire timechain. The `KeyframeSidecar` encodes these identical-value consecutive ranges instead of individual values. As a result, the entire `validation_status` data structure is only a few integers in memory, while still supporting precise chain reorganizations and concurrency protection.

# 7. Future Work

Hornet Node is under active development. While it currently implements header and block validation against mainnet, some aspects are not yet complete, including full script execution, UTXO set evolution, disk storage, and peer management. These will be added in the coming months.

Hornet DSL is an evolving specification that will be hardened once Hornet Node achieves a pure, declarative implementation of all consensus logic. At that point, we anticipate building a compiler for Hornet DSL that transforms the specification into an executable C++ library. Other language backends could follow.

In parallel, we intend to set up a cloud-based application for intelligent consensus correctness testing across clients, as outlined in section 5.4.

We believe that this direction of work will lead us to something Bitcoin has never had: a formal and verifiably correct, self-contained, implementation-neutral, executable specification of Bitcoin consensus -- fully in agreement with, yet distinct from, the reference client.

# 8. Conclusion

Hornet Node demonstrates that Bitcoin's consensus rules can be expressed as a compact, executable specification, whether in idiomatic C++ or in a purpose-built domain-specific language. By separating validation logic from state, and structure from metadata, Hornet offers a clean and modular alternative to legacy implementations. Its focus on clarity, conciseness, and modern expression make it a promising platform for low-level Bitcoin education and experimentation.

While not yet feature complete, Hornet Node already serves as a working example of how Bitcoin consensus can be implemented in a way that is rigorous, readable, and performant. We hope it contributes to ongoing efforts toward client diversity, verifiability, and performance.

## About the Author

Toby Sharp is a mathematician, software developer, and system architect specializing in high-performance real-time C++ systems, including computer vision, augmented reality, and numerical optimization. He was previously Principal Software Scientist at Microsoft (2005-2022) and is now Lead Software Architect at Google (2022-).

His [publications](#) are mostly in realtime computer vision. He has received the MacRobert Gold Medal from the Royal Academy of Engineering (2011), the IEEE Computer Vision Foundation Longuet-Higgins Prize (2021), and the IEEE Computer Vision and Pattern Recognition Best Paper Prize (2011). He is also the lead developer of Microsoft's HoloLens 2 Articulated Hand Tracking (2019), Azure Kinect Body Tracking (2012, 2020), and Kinect Fusion SDK (2015). He now works on Android XR.

Hornet Node is a self-funded spare-time passion project. If you would like to support this work, you may do so at `33Y6TCLKvgjgk69CEAmDPijgXeTaXp8hYd`.

# References

[1] Nakamoto, S. (2008). *Bitcoin: A Peer-to-Peer Electronic Cash System*.
https://bitcoin.org/bitcoin.pdf

[2] Bitcoin Core Developers. *Bitcoin Core*. GitHub.
https://github.com/bitcoin/bitcoin

[3] T. Sharp (2025). *Hornet Node: A Declarative, Executable Specification of Bitcoin Consensus*.
https://hornetnode.org/paper.html

[4] NVK (2025). *BTC242: Bitcoin Core vs Knots w/ NVK*, hosted by Preston Pysh. The Investor Podcast.
https://www.theinvestorspodcast.com/bitcoin-fundamentals/btc242-bitcoin-core-vs-knots-w-nvk/

[5] Luke Dashjr. *Bitcoin Knots*.
https://bitcoinknots.org

[6] C. Moody, *Clark Moody Bitcoin Dashboard*.
https://bitcoin.clarkmoody.com/bitcoin

[7] De Filippi, P., & Loveluck, B. (2015).
*Analyzing the 2013 Bitcoin fork: centralized decision-making saved the day. Princeton CITP Blog.*
https://blog.citp.princeton.edu/2015/07/28/analyzing-the-2013-bitcoin-fork-centralized-decision-making-saved-the-day/

[8] Bartoletti, M., Zunino, R. (2018). *BitML: A Calculus for Bitcoin Smart Contracts*.
https://arxiv.org/abs/1804.07574

[9] Russell O'Connor et al. (2017). *Simplicity: A New Language for Blockchains*.
https://blockstream.com/simplicity.pdf

[10] McKeeman, W.M. (1998). *Differential Testing for Software*. Digital Technical Journal.

[11] Pierce, B. C. et al. (2009). *Software Foundations.*
https://softwarefoundations.cis.upenn.edu/

[12] Chen et al. (2023). *Automated Fuzzing with Large Language Models.*
https://arxiv.org/abs/2302.08530

[13] Gamma, E., Helm, R., Johnson, R., & Vlissides, J. (1994). *Design Patterns: Elements of Reusable Object-Oriented Software*. Addison-Wesley. (Chapter 5: *Visitor*, pp. 331–344)

[14] Bitcoin Core Developers. *libbitcoinkernel*.
https://github.com/bitcoin/bitcoin/tree/master/src/kernel